\documentclass[%
reprint, 
 amsmath,amssymb,
 aps, physrev,
 prl,
]{revtex4-2}

\usepackage{graphicx}
\usepackage{dcolumn}
\usepackage{bm}
\usepackage{comment} 
\usepackage{subcaption} 

\begin{document}

\preprint{APS/123-QED}



\title{Channel-loss-independent quantum-enhanced interferometer} 

\author{Yi-Xin Shen}
\affiliation{State Key Laboratory of Low Dimensional Quantum Physics, Department of Physics, Tsinghua University, Beijing 100084, China}

\author{Zhou-Kai Cao}
\affiliation{State Key Laboratory of Low Dimensional Quantum Physics, Department of Physics, Tsinghua University, Beijing 100084, China}

\author{Jian Leng}%
\affiliation{State Key Laboratory of Low Dimensional Quantum Physics, Department of Physics, Tsinghua University, Beijing 100084, China}

\author{Xiang-Bin Wang}
\email{xbwang@mail.tsinghua.edu.cn}
\affiliation{State Key Laboratory of Low Dimensional Quantum Physics, Department of Physics, Tsinghua University, Beijing 100084, China}


\date{\today}

\begin{abstract}
We propose a channel-loss-independent quantum-enhanced interferometer. In our scheme, the Fisher information for phase difference of weak light from a remote star remains constant under arbitrarily large channel loss, and the angular resolution of our method is better than that of prior quantum-enhanced methods in the long-baseline regime. Moreover, our method requires only threshold detectors and tunable coherent state or two-mode squeezed state sources, both of which are matured technologies nowadays.

\end{abstract}
\maketitle
\section{Introduction}
The optical interferometer is a fundamental technique for detecting distant astronomical objects with ultra-high precision. Traditional optical interferometers rely on transmitting stellar photons through baselines of two telescopes to achieve (post-selected) single-photon interference. In principle, the longer the baseline length, the higher the precision of the measurement result of interference is. However, implementing such single-photon interference with a long baseline is highly challenging, especially because of the inherent difficulties in manipulating photons from stars, which exhibit unstable frequencies, random arrival times, and other unpredictable characteristics. An important breakthrough \cite{gottesman2012longer} is to introduce auxiliary single-photon entanglement to enhance the observation, i.e. quantum enhanced telescope. These schemes no longer require controlling and transmitting the incoming stellar photons; instead, they only need to manipulate single photons generated by controllable sources. This greatly reduces experimental complexity and enables baseline lengths far exceeding those of existing traditional methods. Nevertheless, for a long baseline interference, the auxiliary quantum state suffers from channel loss, and this limits the achievable distance in practice. While quantum repeaters can in principle extend the distance, they require challenging technologies.

Here we propose a channel‐loss–independent (CLI) quantum‐enhanced interferometer, which can work with whatever channel loss. Our CLI method requires neither quantum repeaters nor ideal single‐photon entangled sources. Moreover, it needs yes/no detectors only, rather than photon-number-resolving detectors. This makes it practically feasible when we use methods with multi inputs of auxiliary quantum states \cite{marchese2023large}.
Our method uses practical auxiliary photon sources, and we tune the intensity of these photon sources according to channel transmittance for a given baseline length.
The auxiliary state can be controlled through tuning parameters in the initial physical state.
Numerical simulations show that our method remains efficient against loss for arbitrarily long baselines. Actually, the Fisher Information keeps a satisfactory constant for whatever large baselines.

\section{Auxiliary quantum states}
\begin{figure}
    \centering
    \includegraphics[width=0.7\linewidth]{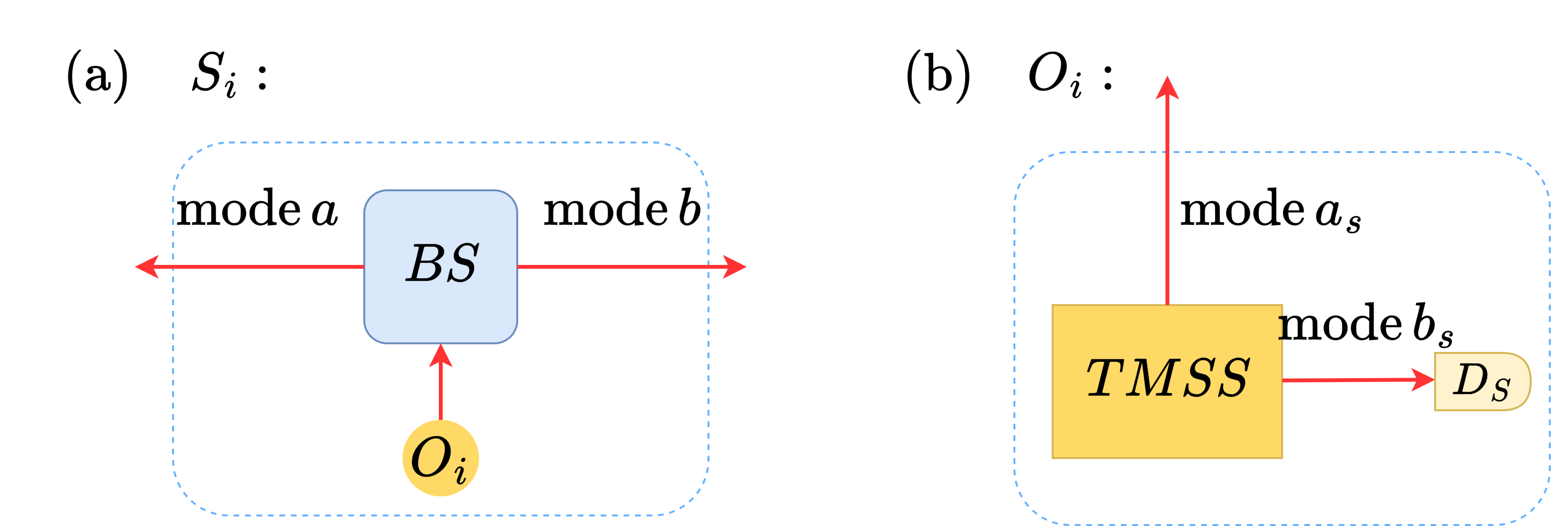}
    \caption{\raggedright Schematic setup of (a) the auxiliary source $S_i$ and (b) the original source $O_i$. T the original source $O_i$ can be either a coherent-state source or a two-mode squeezed state (TMSS) source with threshold detector $D_S$. BS: 50:50 beamsplitter.}
    \label{fig:Fig0}
\end{figure}
As shown in Fig. \ref{fig:Fig0} (a), we define the combination of the original physical source $O_i$ and the 50:50 beamsplitter as the auxiliary source $S_i$ , and the resulting two-mode state of the beams \textit{arriving at the telescopes} as the auxiliary state.
The input beam of the BS comes from the original source $O_i$, which may be a simple single-mode source or a heralded two-mode source — e.g., a two-mode squeezed state (TMSS) with heralding (Fig. \ref{fig:Fig0}(b)). In the heralded case, heralding of mode $b_s$ by clicking on detector $D_S$ post-selects mode $a_s$ as the auxiliary input.

\begin{figure}
    \centering
    \includegraphics[width=0.7\linewidth]{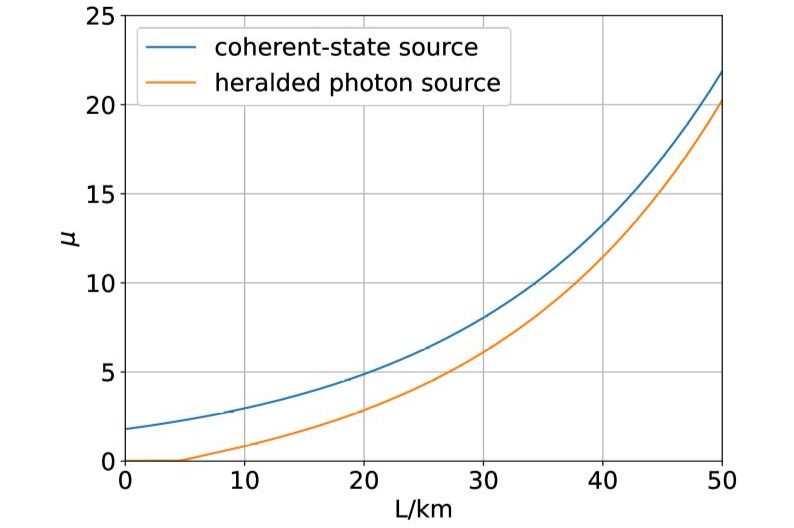}
    \caption{\raggedright The optimized parameter values $\mu$ of the initial physical sources in the calculation of Fisher Information in Fig. \ref{fig:2.5}.}
    \label{fig:3new}
\end{figure}

Our auxiliary photon source is channel-loss-independent (CLI): by optimizing the intensity of the physical source in $O_i$ under any channel transmittance $\eta$, the probability of the single-photon entangled state in the transmitted mixed state remains constant.
As shown in Sec. III, the final result of our quantum-enhanced interferometer is almost independent of channel loss if we use the optimized source parameters according to the specific channel loss. 
The optimized parameter values of methods with single auxiliary sources are shown in Fig. \ref{fig:3new}.

\subsection{Coherent state of source $O_i$}
We can generate the two-mode auxiliary state by directly using a coherent photon source:
\begin{equation}\label{equ:state_coherent}
    |\sqrt{\mu}\mathrm{e}^{i\delta}\rangle=\sum_n\mathrm{e}^{-\frac{\mu}{2}}\frac{\mu^{\frac{n}{2}}\mathrm{e}^{i n\delta}}{\sqrt{n!}}|n\rangle.
\end{equation}
Since our method is supposed to use simple and mature techniques only, we do not take control or manipulations to the global phase $\delta$ here and hence the value $\delta$ may drift randomly. We use the following mixed state for coherent states:
\begin{equation}\label{equ:rho_coherent}
    \rho_C=\sum_{n=0}^{\infty}\mathrm{e}^{-\mu}\frac{\mu^n}{n!}|n\rangle\langle n|.
\end{equation}
Suppose the channel transmittance from BS to a telescope is $\eta$ (this means the channel transmittance for beam $a$ or $b$ is $\eta$).
In arriving at telescopes, our auxiliary quantum state is
\begin{align}\label{equ:rho_ab_coherent}
        \rho_{ab}&=\sum_{m=0}^{\infty}\mathcal{P}_m|\psi_m\rangle\langle\psi_m|,
\end{align}
where
\begin{equation}\label{equ:p_ab_coherent}
    \mathcal{P}_m=\mathrm{e}^{-\eta\mu}\frac{\left(\eta\mu\right)^m}{m!},
\end{equation}
and
\begin{equation}\label{equ:psi_n}
|\psi_m\rangle=\frac{(a_i^{\dagger}+b_i^{\dagger})^m}{\sqrt{2^m m!}}|\text{vac}\rangle,
\end{equation}
for example, $|\psi_0\rangle=|0\rangle|0\rangle$ and $|\psi_1\rangle=\frac{|0\rangle|1\rangle+|1\rangle|0\rangle}{\sqrt{2}}$.
As shown later in Sec. III, the final result of our quantum-enhanced interferometer is almost independent of channel loss if we use the optimized source parameters according to the specific channel loss.

Remark: We don’t worry about whether the global phase $\delta$ in Eq. \eqref{equ:state_coherent} is perfectly random. As shown later, our major result is simply independent of the value of the global phase $\delta$, no matter if it is random or not random, or taking whatever time-dependent values.

\subsection{Heralded source $O_i$ from TMSS}

We can generate the auxiliary quantum state through using a two-mode initial state with heralding as shown in Fig. \ref{fig:Fig0} (b): we choose mode $a_s$ conditional on the heralding of mode $b_s$, i.e., clicking of detector $D_S$.
The state of the TMSS is: 
\begin{equation}\label{equ:heraldstate}
    |\Psi\rangle_{H}=\sqrt{1-r^2}\sum_n r^n|n\rangle|n\rangle=\sum_n\sqrt{p_n}|n\rangle|n\rangle,
\end{equation}
where $p_n=(1-r^2)r^{2k}=\frac{\mu^n}{(1+\mu)^{n+1}}$, and
\begin{equation}\label{equ:mu_defi}
    \mu= \frac{r^2}{1-r^2}
\end{equation}
denotes the mean photon number of the source.
Given ideal detector $D_S$, the auxiliary quantum state is
\begin{align}\label{equ:rho_3_2}
\rho_{ab}&=\sum_{m=0}^{\infty}\mathcal{P}_m|\psi_m\rangle\langle\psi_m|,
\end{align}
with
\begin{equation}\label{equ:prob_2}
    \begin{aligned}
        \mathcal{P}_0 &= \frac{1-\eta}{1+\eta\mu},\\
        \mathcal{P}_m &= \frac{1+\mu}{\mu} \frac{(\mu \eta)^m}{(1+\mu \eta)^{m+1}}.
    \end{aligned}
\end{equation}
According to Eq. \eqref{equ:mu_defi}, here $\mu$ is determined by the tunable parameter $r$ of the initial physical source.

\section{Channel-loss-independent interferometer over large baseline}
\begin{figure}
    \centering
    \includegraphics[width=0.7\linewidth]{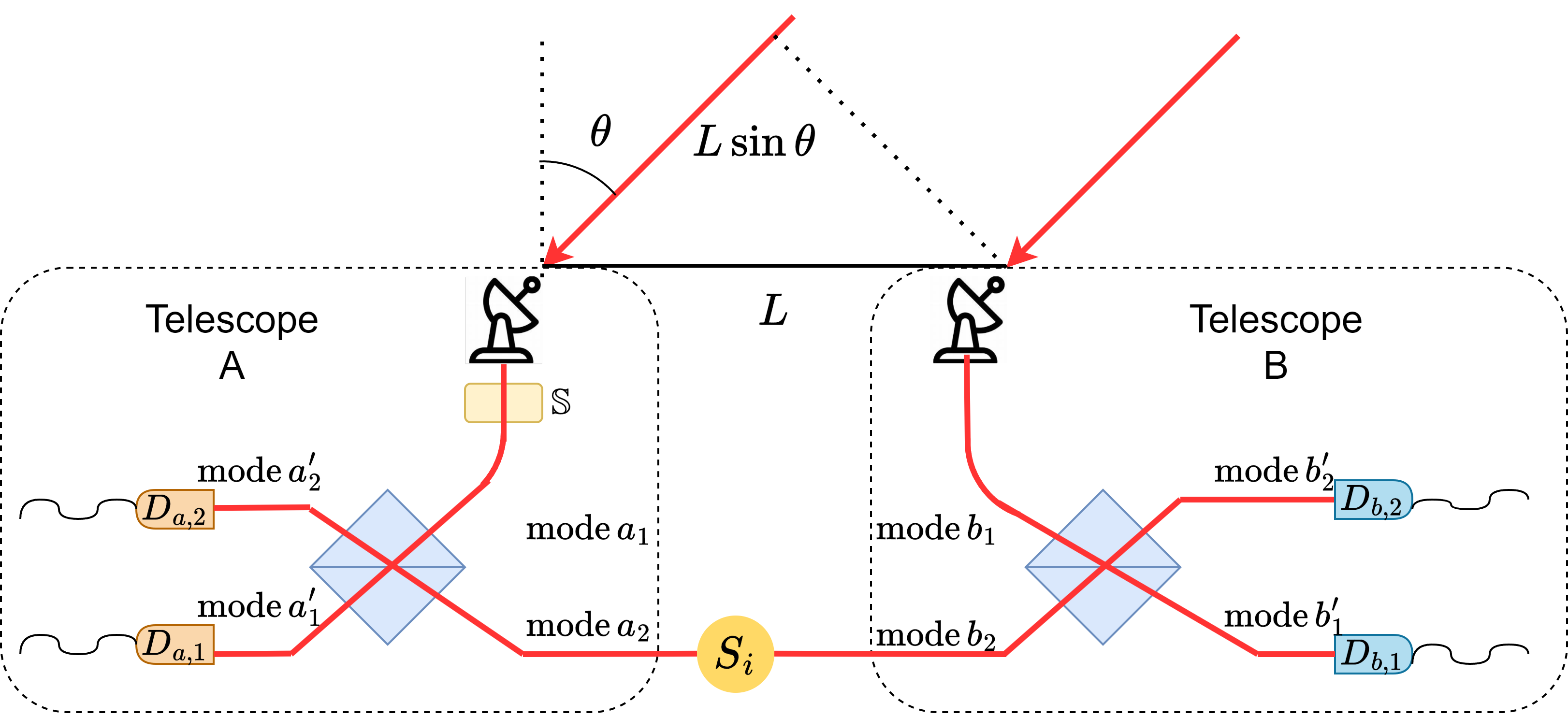}
    \caption{\raggedright Schematic setup of the CLI quantum-enhanced interferometer method with one auxiliary photon source: The two telescopes are separated by a distance $L$, with each telescope independently equipped with a 50:50 beamsplitter and 2 threshold detectors. The auxiliary photon source is deployed at the midpoint between the two telescopes.}
    \label{fig:2}
\end{figure}

\begin{figure}
    \centering
    \includegraphics[width=0.7\linewidth]{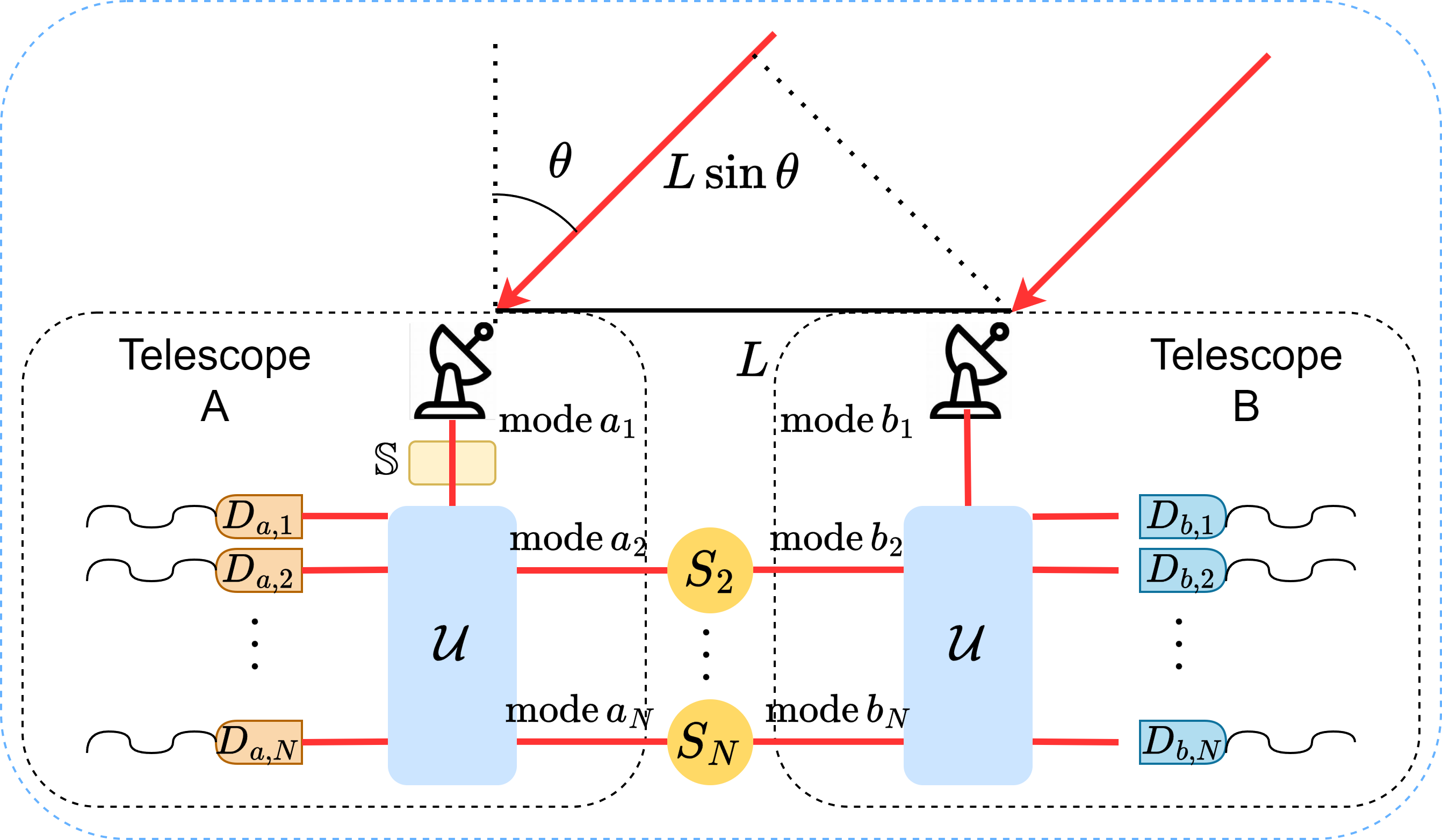}
    \caption{\raggedright Schematic setup of our method with multi auxiliary photon sources. Each telescope is equipped with a linear optical network $U$ and $N$ threshold detectors.}
    \label{fig:Fig1}
\end{figure}

We propose the setup of our Channel-loss-independent quantum-enhanced interferometer with a single auxiliary source and multi auxiliary sources in Fig. \ref{fig:2} and Fig. \ref{fig:Fig1}, respectively.
Each auxiliary source is connected to the linear-optical circuits at the two telescope stations via a pair of equal-length optical fibers. Note that, different from the prior works \cite{gottesman2012longer, marchese2023large}, here in our setup, we have proposed to replace the perfect single photon sources with practical sources, which rely solely on matured technologies, and whose intensities are tunable. This makes the base of our channel-loss-independent quantum-enhanced telescope: by choosing the intensity value according to the baseline loss, our setup can work efficiently with arbitrarily large channel loss of the baseline.
Under the framework of \cite{gottesman2012longer, marchese2023large}, we also use linear optical circuits for our method. In the method with multi auxiliary photon sources, each telescope is equipped with a $N\times N$ linear optical circuit consisting of beamsplitters and phase shifters only and $N$ threshold detectors. $N-1$ auxiliary sources are placed at the midpoint of the baseline between the two telescopes.

\subsection{Received quantum states of telescopes from star}
For the target source at angular position $\theta$ in Fig. \ref{fig:Fig1}, the emitted light attenuates into a mixture of vacuum state and single‐photon entangled state by the time it reaches the telescopes:
\begin{equation}\label{equ:01}
    |\varphi_1\rangle=\frac{\mathrm{e}^{i\phi}\hat{a}_1^{\dagger}+\hat{b}_1^{\dagger}}{\sqrt{2}}|vac\rangle,\quad\phi\equiv\frac{2\pi L\sin\theta}{\lambda},
\end{equation}
where $\lambda$ is the wavelength, $a_1$ and $b_1$ are the modes of telescopes A and B, respectively. The two telescopes are separated by a distance $L$ (the baseline length). We denote the mean photon number of the starlight received by the telescopes as $\epsilon$. Consequently, in a fraction $\epsilon$ of the time windows, the two telescopes share the single‐photon entangled state $\lvert\varphi_1\rangle$ defined in Eq. \eqref{equ:01}; in the remaining windows, both telescopes receive the vacuum state.

\subsection{The method with single auxiliary photon source}
Here we demonstrate how the CLI method with one auxiliary photon source determines the value of $\phi$ defined in Eq. \eqref{equ:01}, and compare the Fisher Information between our CLI method and the method with auxiliary single-photon source \cite{gottesman2012longer}.

\subsubsection{Our method.}
The setup of the method with one auxiliary heralded photon source is shown in Fig. \ref{fig:2}. We use threshold detectors and
we count the frequency of each kind of events defined by the joint click of the detectors $\{D_{a,1}, D_{a,2}, D_{b,1}, D_{b,2}\}$. In particular, 
we denote the events of click outcomes by $\vec{m}=(b_{a,1}, b_{a,2},b_{b,1},b_{b,2})$, where $b_{x,i}=\mathrm{bool}(n_{x,i})\;(x\in\{a,b\}, i\in\{1,2\})$ means $b_{x,i}=Y$ when the threshold detector $D_{x,i}$ clicks and $b_{x,i}=N$ when the threshold detector $D_{x,i}$ is silent. 

Since we choose optimized parameter values for the source according to the channel loss of each channel distance, the performance of our method is channel-loss-independent. Moreover, we need some other values to complete the calculation of phase difference $\phi$ and Fisher Information.
We need the probability of different types of events $\{\vec{m}\}$. In a real experiment, we also need to record the number of each kind of events $\{\vec{m}\}$ through observing those threshold detectors.
Based on this, we have the constraints for value $\phi$ and $P(\vec{m}|\phi)$. Given these constraints, we can calculate the $\phi$ value through MLE. However, in the calculation of Fisher Information alone, we only need the theoretical value of $P(\vec{m}|\phi)$ which can be calculated by Eqs. \eqref{equ:Pd_propto2}-\eqref{equ:Pd_propto3} directly, and we do not need the (simulation of) experimentally observed data though the simulation is needed in calculating $\phi$ by MLE. Below, we derive these constraints.
The actual starlight is often very weak, so it is a mixture of lots of vacuum and single-photon states as shown by Eq. \eqref{equ:01}. Mathematically, the density operator of starlight is
\begin{equation}
    \rho_S=(1-\epsilon)|vac\rangle\langle vac|+\epsilon|\varphi_1\rangle\langle\varphi_1|
\end{equation}
where $\epsilon$ is a small positive number and $|\varphi_1\rangle$ is defined by Eq. \eqref{equ:01}.

Given the interference of the starlight and the auxiliary light, we can compute the probability $P(\vec{m}|\phi)$. Asymptotically,
\begin{equation}\label{equ:Pm_specific}
    P(\vec{m}|\phi)=\sum_{\{\vec{d}|\vec{d}\in S_{\vec{m}}\}}\epsilon\left(\mathcal{A}_{\vec{d}}+\mathcal{B}_{\vec{d}}\cos\phi\right)+\mathcal{C}_{\vec{d}}.
\end{equation}
We will present the detailed derivation of Eq. \eqref{equ:Pm_specific} in Sec.III B 3: “Value of $P(\vec{m}\mid\phi)$.”

\subsubsection{Fisher Information.}
We quantify the performance of methods with different auxiliary sources by comparing their Fisher information.
The Fisher information is defined as
\begin{equation}\label{equ:Fisher_Info}
    \mathcal{F}(\phi)=\sum_{\vec{m}}\left(\frac{\partial P\left(\vec{m}|\phi\right)}{\partial \phi}\right)^2\frac{1}{P\left(\vec{m}|\phi\right)}.
\end{equation}
The variance of our method, $(\delta\phi)^2$, is proportional to the inverse of the Fisher information:
\begin{equation}
    (\delta\phi)^2\geq\frac{1}{N_t\mathcal{F}(\phi)},
\end{equation}
where $N_t$ is the number of independent experiments. The corresponding angular uncertainty $\delta\theta$ is
\begin{equation}
    \delta\theta=\frac{\lambda\delta\phi}{2\pi L}\geq\frac{\lambda}{2\pi L}\frac{1}{\sqrt{N_t\mathcal{F}(\phi)}}.
\end{equation}


\subsubsection{Value of $P(\vec{m}|\phi)$.}
 To calculate the Fisher Information above, we need the value of  $P(\vec{m}|\phi)$ as defined around Eq. \eqref{equ:Pm_specific}. In our calculation of the Fisher Information, we only need the asymptotic value of $P(\vec{m}|\phi)$.
For presentation convenience, we introduce a virtual setup where we use photon number resolving detectors, say, what would happen if we had photon number resolving detectors. We represent events in this imaginary setup by
$
\vec{d} = \left(n_{a,1},\,n_{a,2},n_{b,1},\,n_{b,2}\right),
$
where $n_{x,i}$ is the number of photons received by detector $D_{x,i}$. For example, notation $\vec{d}=(1,0,1,0)$ represents the state that one photon arrives at detector $D_{a,1}$ and one photon arrives at detector $D_{b,1}$, while detectors $D_{a,2}$ and $D_{b,2}$ receive no photon.
We use definition $\vec{d}$ for the purpose of notational simplicity only. We emphasize that our method does not use any photon-number resolving detector as actual detection is performed using yes/no detectors.

Theoretically, the value of phase difference $\phi$ is determined by $P(\vec{m}|\phi)$, the probability of finding each event $\vec{m}$ in the real setup is
\begin{equation}\label{equ:Pd_propto2}
    P(\vec{m}|\phi)=\sum_{\{\vec{d}|\vec{d}\in S_{\vec{m}}\}}P(\vec{d}).
\end{equation}
With this formula, to complete the calculation formulas of Fisher Information, we only need the calculation formula $P(\vec{d})$ as shown below
\begin{equation}\label{equ:Pd_propto3}
    P(\vec{d})=\epsilon\left(\mathcal{A}_{\vec{d}}+\mathcal{B}_{\vec{d}}\cos\phi\right)+\mathcal{C}_{\vec{d}}.
\end{equation}
The calculation of the coefficients $\mathcal{A}_{\vec{d}}$, $\mathcal{B}_{\vec{d}}$ and $\mathcal{C}_{\vec{d}}$ can be seen in Eqs.(S8)-(S11) of the Supplemental Material \cite{SM}.

\subsubsection{Numerical results of Fisher Information.}
\begin{figure}
    \centering
    \begin{minipage}[t]{0.48\linewidth}
        \centering
        \includegraphics[width=\linewidth]{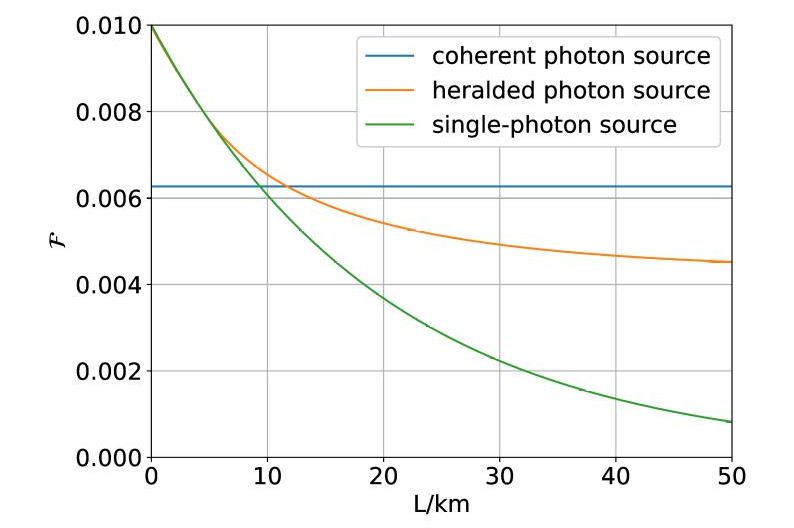}
        \caption*{(a)}
    \end{minipage}
    \begin{minipage}[t]{0.48\linewidth}
        \centering
        \includegraphics[width=\linewidth]{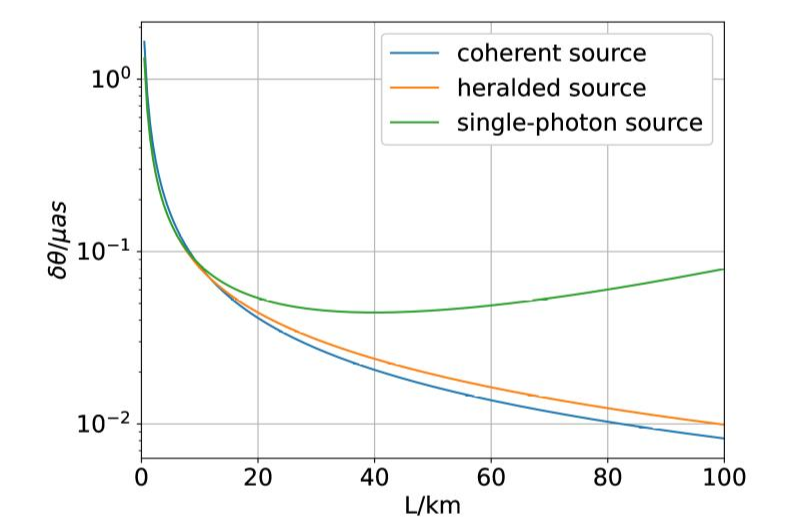}
        \caption*{(b)}
    \end{minipage}
    \caption{\raggedright Performance comparison of different methods using (a) Fisher information and (b) the angular uncertainty under different baseline lengths. We take $N_t = 10^5$ independent measurements in (b). The comparison is conducted under a fixed starlight intensity of $\epsilon = 0.02$. We have taken the optimized value $\mu$ in the initial physics source for the calculation of each distance point, as shown in Fig. \ref{fig:3new} of the End Matter.}
    \label{fig:2.5}
\end{figure}

 Using Eq. \eqref{equ:Fisher_Info} together with Eqs. \eqref{equ:Pd_propto2}-\eqref{equ:Pd_propto3}, the Fisher Information of our method is numerically calculated. Fig. \ref{fig:2.5}(a) shows the optimal Fisher information of methods using a heralded or a coherent photon source at various baseline lengths $L$. These methods are compared with the existing single-photon source method \cite{gottesman2012longer}.
By tuning the source intensity to maximize $\mathcal{P}_1$, both the heralded and coherent source methods maintain nearly constant Fisher information under low channel transmittance, demonstrating the channel-loss-independent nature of the approach.
Fig. \ref{fig:2.5}(b) further compares the optimal angular uncertainties of different methods for $N_t = 10^5$ independent experiments, based on the Fisher information shown in Fig. \ref{fig:2.5}(a). While the method using a single-photon source reaches its resolution limit at a baseline length of $40\,\mathrm{km}$, our methods continue to improve in angular resolution as the baseline length increases. 

\subsubsection{Advantages of heralded source.}
1.Value of $\epsilon$ in the density operator of starlight. In the calculation above, we assumed a known value for $\epsilon$. However, determination of this value itself is not so trivial. The starlight can be very weak, and its state does not have to be stable. Therefore, it is not convincing to simply take an additional experiment to find the value $\epsilon$ before or after the interference experiment with auxiliary quantum states because the (average) value $\langle\epsilon\rangle$ can be different for the above two experiments done at different times. To convincingly estimate the value $\epsilon$ in an experiment, we can randomly choose some time windows to send out vacuum auxiliary states. If we use this method, the way of using a heralded auxiliary source seems to be more convenient: it does not need any active operation because it can simply use those non-heralded states of the beam $a_s$.

2. Higher performance with multi-inputs auxiliary heralded photon sources. As has been shown numerically already, the coherent state auxiliary source has a higher Fisher Information than that of the heralded photon source if we only use one auxiliary photon source. However, the result of the coherent auxiliary source has no improvement even if we use multi-inputs, while the result is significantly improved if we use multi-input auxiliary heralded photon sources, as shown in Sec. III C.

\subsubsection{Results of MLE.}
Although Fisher Information represents the theoretically achievable result of our method, in a real experiment, we can only compute the value $\phi$ by a specific numerical method such as MLE.
Experimentally, we record the counts of different outcomes $\vec{m}$, denoted by $\hat{N}_{\vec{m}}$.
We set the phase shifter $\mathbb{S}$ in Fig. \ref{fig:Fig1} to two different values, $\mathbb{S}=0$ and $\mathbb{S}=\pi/2$, and for each phase-shifter configuration, we take $N_t$ independent experiments. The experimentally observed counts of events $\{\vec{m}\}$ with probabilities $\{P\left(\vec{m}|\phi\right)\}$ (zero phase shift on $\mathbb{S}$) are denoted as $\{\hat{N}_{\vec{m}}\}$, and the counts of events with probabilities $\left\{P\left(\vec{m}|\phi + \pi/2\right)\right\}$ ($\frac{\pi}{2}$ phase shift on $\mathbb{S}$) are denoted as $\{\hat{N}^\prime_{\vec{m}}\}$.
The negative log-likelihood function is defined as:
\begin{equation}\label{equ:MLE2}
    \mathcal{L}(\phi)=-\sum_{\{\vec{m}\}}\left( \hat{N}_{\vec{m}}\log\left(P(\vec{m}|\phi)\right)+\hat{N}^\prime_{\vec{m}}\log\left(P(\vec{m}|\phi+\pi/2)\right)\right).
\end{equation}
The estimated value $\tilde{\phi}$ is
\begin{equation}
    \tilde{\phi} =\arg\min_{\phi}\bigl[\mathcal{L}(\phi)].
\end{equation}
The estimated angle value $\tilde{\theta}$ is then
\begin{equation}
    \tilde{\theta}=\arcsin\left(\frac{\lambda\tilde{\phi}}{2\pi L}\right).
\end{equation}

\begin{figure}
    \centering
    \includegraphics[width=0.7\linewidth]{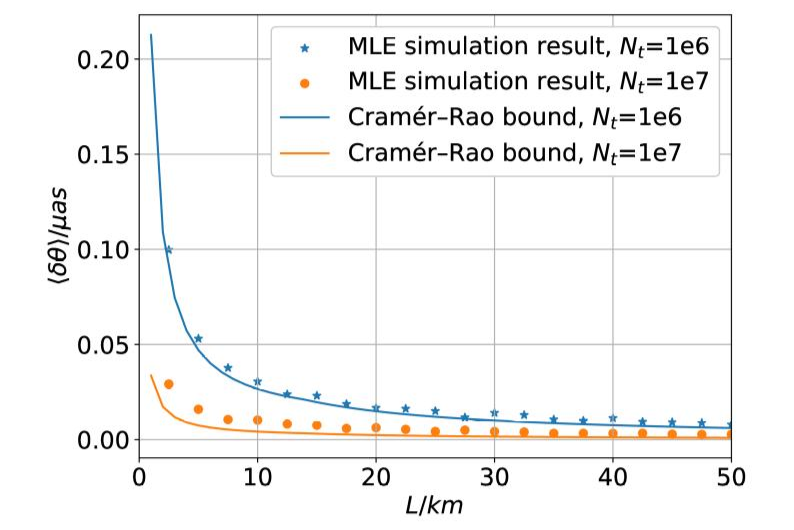}
    \caption{\raggedright Angular uncertainty of our method and the Cram\'{e}r-Rao bound. $N_t:$ the number of effective time windows for each phase shift on $\mathbb{S}$. For each simulation data point, hundreds of random angle values $\theta$ uniformly distributed in $[0,\lambda/L]$ are generated, and all possible measurement outcomes are simulated to compute the average deviation between the estimated and true angle values.}
    \label{fig:enter-label3}
\end{figure}

We numerically simulate the average angular uncertainty $\langle\delta\theta\rangle$ between the exact value $\theta$ and the estimated value $\tilde{\theta}$ for the method with a single heralded source. The numerical results are shown in Fig~\ref{fig:enter-label3}. 
The simulation result is close to the Cram\'{e}r-Rao bound predicted by the Fisher Information of the system, indicating that our method with MLE maintains high precision under high channel loss.

\subsection{Our method with multi auxiliary photon sources}
As shown by \cite{marchese2023large}, the performance of a quantum-enhanced interferometer can be improved using multi-inputs of single-photon entanglement sources. Here we use multi-inputs of heralded photon sources to improve the Fisher Information of our method. Different from the prior artwork \cite{marchese2023large}, our method requires neither single-photon sources nor photon number-resolving detectors.

Recall the setup in Fig. \ref{fig:Fig1}, each telescope is attached with a $N\times N$ linear optical circuit $U$, whose matrix elements are $U_{i,j}=\frac{\exp\bigl[(i-1)\,(j-1)\,\pi/N\bigr]}{\sqrt{N}}$. The $U$ circuit can be implemented with a linear-optical circuit containing at most $\frac{N(N-1)}{2}$ beam splitters and phase shifters \cite{reck1994experimental, clements2016optimal}, and hence has no significant challenges under current technologies.

The $N$ output modes of the linear optical circuit $U$ are each measured by one threshold detector.
We denote the outcomes of detectors $\{D_{a,1},\cdots D_{a,n},D_{b,1},\cdots D_{b,n}\}$ by the vector $\vec{m}=(b_{a,1},\cdots b_{a,N},b_{b,1},\cdots b_{b,N}\}$.
The probability of each outcome $\vec{m}$ is
\begin{equation}\label{equ:Pm_specific2}
    P(\vec{m}|\phi)=\sum_{\{\vec{d}|\vec{d}\in S_{\vec{m}}\}}\epsilon\left(\mathcal{A}^{\prime}_{\vec{d}}+\mathcal{B}^{\prime}_{\vec{d}}\cos\phi\right)+\mathcal{C}^{\prime}_{\vec{d}},
\end{equation}
which is in the same form as Eq. \eqref{equ:Pm_specific}, where $S_{\vec{m}}$ is the set for all events $\{\vec{d}\}$ of the virtual method that can cause the measurement outcome of $\vec{m}$. Given Eq. \eqref{equ:Pm_specific2}, we can quantify the performance of our method with the Fisher information $\mathcal{F}(\phi)$ defined in Eq. \eqref{equ:Fisher_Info}.

\begin{figure}
    \centering
    \includegraphics[width=0.7\linewidth]{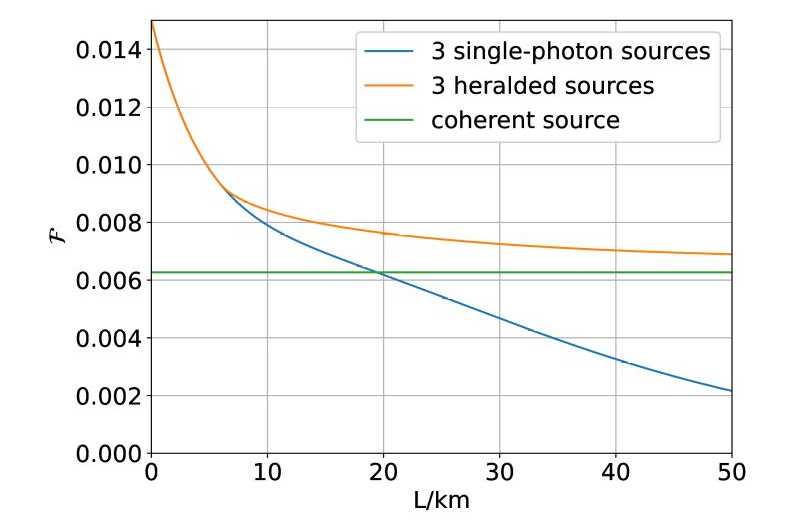}
    \caption{\raggedright Performance comparison of different methods using Fisher information. The comparison is conducted under a fixed starlight intensity of $\epsilon = 0.02$.}
    
    \label{fig:enter-label2}
\end{figure}

\textit{Numerical results}. We have also compared the Fisher Information of our method with 3 auxiliary heralded sources and the existing method with 3 ideal single-photon sources under different baseline lengths; the results are shown in Fig. \ref{fig:enter-label2}. Evidently, the performance of our method with multiple auxiliary heralded sources behaves better than the method with a single auxiliary coherent source and the existing method with 3 ideal single-photon sources \cite{marchese2023large} under low transmittance. 

\section{Conclusion}
We demonstrate the channel-loss-independent quantum-enhanced interferometer method and show its advantage over existing quantum-enhanced results \cite{gottesman2012longer, marchese2023large, tsang2011quantum, gottesman2012longer, khabiboulline2019optical, howard2019optimal, lupo2020quantum, huang2021quantum, huang2022imaging, zanforlin2022optical, marchese2023large, brown2023interferometric, huang2023quantum, wang2021superresolution,khabiboulline2019quantum, bland2021quantum, bojer2022quantitative, stankus2020two,chen2023astrometry}. Our CLI quantum-enhanced interferometer is experimentally feasible, requiring only threshold detectors, coherent-state sources or two-mode squeezed photon sources, beamsplitters, and phase shifters. By tuning the intensity of the initial physical sources, the baseline can be made arbitrarily long while the Fisher information keeps almost constant.

\begin{acknowledgements}
    We acknowledge the financial support in part by National Natural Science Foundation of China grant No.12174215 and No.12374473. This study is also supported by the Taishan Scholars Program. 
\end{acknowledgements}

\nocite{*}
\bibliography{apssamp.bib}

\newpage
\section*{Supplemental Material for ``Channel-Loss-Independent Quantum-Enhanced Interferometer"}
\subsection{A More Detailed Introduction to the Auxiliary Photon Sources}
The setup of our auxiliary source $S_i$ and the original source $O_i$ are shown in Fig. \ref{fig:Fig0}. The physical source in $O_i$ can be a two-mode squeezed-state source (TMSS) or other types of physical sources.
In general, we consider the following mathematical form of the heralded source state of beam $a_S$ under the heralding condition of beam $b_S$ (the condition that detector $D_S$ clicks)
\begin{align}\label{equ:rho_0}
    \rho_0=\sum_{n=0}^\infty p_n|n\rangle\langle n|,
\end{align}
where $|n\rangle$ is the Fock state. After passing through the 50:50 BS, we obtain the new two-mode state
\begin{align}\label{equ:rho_1}
    \rho_1=\sum_{n=0}^\infty p_n|\psi_n\rangle\langle \psi_n|,  
\end{align}
where
\begin{equation}\label{equ:psi_n2}
|\psi_n\rangle=\frac{(a_i^{\dagger}+b_i^{\dagger})^n}{\sqrt{2^n n!}}|\text{vac}\rangle,
\end{equation}
for example, $|\psi_0\rangle=|0\rangle|0\rangle$ and $|\psi_1\rangle=\frac{|0\rangle|1\rangle+|1\rangle|0\rangle}{\sqrt{2}}$.
Modes $a_i$ and $b_i$ are outputs of BS.
For convenience, we shall regard the state of the beams arriving at the telescopes as our ancillary quantum state.
Suppose the channel transmittance from beamsplitter to a telescope is $\eta$ (this means the channel transmittance for beam $a$ or $b$ is $\eta$).
In arriving at telescopes, our ancillary quantum state is
\begin{align}\label{equ:rho_3}
        \rho_{2}&=\sum_{m=0}^{\infty}\mathcal{P}_m|\psi_m\rangle\langle\psi_m|,
\end{align}
where
\begin{equation}\label{equ:4}
    \mathcal{P}_m=\eta^m \sum_{n=m}^{\infty} p_n \binom{n}{m} (1-\eta)^{n-m}.
\end{equation}
We can calculate the auxiliary state of different physical sources with Eqs. \eqref{equ:rho_3}-\eqref{equ:4}.

\subsection{The detailed derivation of probabilities of the virtual method.}
\subsubsection{The virtual method with a single auxiliary source}
For any auxiliary source state in the form of Eqs. \eqref{equ:rho_3}-\eqref{equ:4}, the state of light beams arriving at detectors $\{D_{a_1}, D_{a_2}, D_{b_1},D_{b_2}\}$ is a mixed state.
\begin{equation}\label{equ:herald_source_detector_state}
    \begin{aligned}
        &\rho_{3}=\sum_{m=0}^{\infty}\mathcal{P}_m\left(\epsilon|\xi_{m}\rangle\langle\xi_{m}|+(1-\epsilon)|\xi_m^\prime\rangle\langle\xi_m^\prime|)\right),\\
    \end{aligned}
\end{equation}
where
\begin{equation}\label{equ:herald_source_detector_state2}
    \begin{aligned}
        &|\xi_{m}\rangle=\frac{1}{\sqrt{m!}}\left(\frac{\hat{a}_0^\dagger-\hat{a}_1^\dagger+\hat{b}_0^\dagger-\hat{b}_1^\dagger}{2}\right)^m|vac\rangle;\\
        &|\xi_{m}^\prime\rangle=\frac{1}{\sqrt{m!}}\left[\frac{\hat{a}_0^\dagger+\hat{a}_1^\dagger+\mathrm{e}^{i\phi}\left(\hat{b}_0^\dagger+\hat{b}_1^\dagger\right)}{2}\left(\frac{\hat{a}_0^\dagger-\hat{a}_1^\dagger+\hat{b}_0^\dagger-\hat{b}_1^\dagger}{2}\right)^m\right]|vac\rangle.
    \end{aligned}
\end{equation}
Given the virtual method with photon-number-resolving detectors introduced in Sec. III of the main text, the probability of observing result $\vec{d}$ of the virtual method is denoted as $P(\vec{d})$.
For the mixed state $\rho_3$ shown in Eq.\eqref{equ:herald_source_detector_state}, the probability of the result $\vec{d}=(n_{a,0},n_{a,1},n_{b,0},n_{b,1})$ of the virtual method is
\begin{equation}\label{equ:Pd_propto}
    \begin{aligned}
        P\left(\vec{d}\right)=&\left|\left\langle\vec{d}\right|\rho_{3}\left|\vec{d}\right\rangle\right|\\=&\sum_{m=1}^{\infty}\left[\mathcal{P}_{m}\left(\epsilon|\langle\vec{d}|\xi_{m}\rangle|^2\right)+\mathcal{P}_{m-1}\left(\left(1-\epsilon\right)|\langle\vec{d}|\xi_{m-1}^\prime\rangle|^2\right)\right]
        \\=&\epsilon\left(\mathcal{A}_{\vec{d}}+\mathcal{B}_{\vec{d}}\cos\phi\right)+\mathcal{C}_{\vec{d}},
    \end{aligned}
\end{equation}
where $\vec{d}=\left(n_{a,0}, n_{a,1}, n_{b,0}, n_{b,1}\right)$ and $n_{a,0}+n_{a,1}+n_{b,0}+n_{b,1}=m$. We denote $\left\langle\vec{d}|\xi^\prime_{m}\right\rangle=A_{m}+B_{m}\mathrm{e}^{i\phi}$, $\left\langle \vec{d}|\xi_{m}\right\rangle=C_m$. The values of the coefficients $A_{m},B_{m}$ and $C_{m}$ can be easily calculated by Eq. \eqref{equ:herald_source_detector_state2}.
For the simplest virtual outcome $\vec{d}$ that only two detectors $D_{a,k}$ and $D_{b,l}$ click, the coefficients are given by
\begin{equation}\label{equ:heralded_result_11}
    \begin{aligned}
        A_m&=\sqrt{\frac{(n)!(m+1-n)!}{2^{m+1}}}U_{1l}U_{2k}^{n}U_{2l}^{m-n-1}\binom{m}{n},\\
   B_m&=\sqrt{\frac{(n)!(m+1-n)!}{2^{m+1}}}U_{1k}U_{2k}^{n-1}U_{2l}^{m-n}\binom{m}{n-1},\\
   C_m&=\sqrt{\frac{(n)!(m-n)!}{2^{m}m!}}U_{2k}^{n}U_{2l}^{m-n}\binom{m}{n},
    \end{aligned}
\end{equation}

Given Eqs. \eqref{equ:Pd_propto}-\eqref{equ:heralded_result_11}, the values of $\mathcal{A}_m,\mathcal{B}_m\; \text{and}\;\mathcal{C}_m$ in Eq.\eqref{equ:Pd_propto} are
\begin{equation}\label{equ:herald_result_important}
    \begin{aligned}
        \mathcal{A}_{\vec{d}}&= \mathcal{P}_{m-1}\epsilon\left(|A_{m-1}|^2+|B_{m-1}|^2\right)-\mathcal{P}_m\left(1-\epsilon\right)|C_m|^2,\\
        \mathcal{B}_{\vec{d}}&= 2\mathcal{P}_{m-1}\epsilon|A_{m-1}||B_{m-1}|,\\
        \mathcal{C}_{\vec{d}}&= \mathcal{P}_m\left(1-\epsilon\right)|C_m|^2,
    \end{aligned}
\end{equation}
Given Eqs. \eqref{equ:Pd_propto}-\eqref{equ:herald_result_important}, all values of $P(\vec{d})$ in Eq.(17) of the main text can be numerically calculated.

Remark: Eq.\eqref{equ:Pd_propto} is valid for any ancillary photon source satisfying Eqs. \eqref{equ:rho_3}–\eqref{equ:4}, but the specific values of $\{\mathcal{A}_m,\mathcal{B}_m,\mathcal{C}_m\}$ depend on the photon-number distribution $\{P_m\}$ of the auxiliary sources.

\subsubsection{The virtual method with multi-inputs auxiliary sources}
For the method with multi-input ancillary heralded photon sources,
the mixed state arrived at the detectors is
\begin{equation}\label{equ:rho_3_multi}
    \rho_{3}=\sum_{\{m\}}\mathcal{P}_{\{m\}}\left(\epsilon|\xi_{\{m\}}\rangle\langle\xi_{\{m\}}|+(1-\epsilon)|\xi_{\{m\}}^\prime\rangle\langle\xi_{\{m\}}^\prime|\right),
\end{equation}
where $\{m\}=\sum_{m_i}\{m_2,\cdots,m_N|m_2+\cdots+m_N=m\}$ denotes the set of source events that the total number of photons is $m$, and
\begin{equation}
    \mathcal{P}_{\{m\}}=\prod_{i=2}^{N}\mathcal{P}_{m_i}=\left(\frac{\mu+1}{\mu}\right)^N\prod_{i=2}^{N}\frac{\left(\eta\mu\right)^{m_i}}{\left(1+\eta\mu\right)^{\left(m_i+1\right)}}
\end{equation}
is the probability of getting source event $\{m\}$;
\begin{equation}
    \begin{aligned}
        &|\xi_{m}\rangle=U^{\otimes2}\left(|vac\rangle\bigotimes_{\{m_i\}}|\psi_{m_i}\rangle\right);\\
        &|\xi_{m}^\prime\rangle=U^{\otimes2}\left(|\varphi_1\rangle\bigotimes_{\{m_i\}}|\psi_{m_i}\rangle\right),
    \end{aligned}
\end{equation}
where $U$ is the $N\times N$ linear optical circuit composed of beamsplitters and phase shifters.
The probability of getting each result $\vec{d}$ of the virtual setup is
\begin{equation}\label{equ:heralded_result_1}
    \begin{aligned}
&P\left(\vec{d}\right)=\left|\left\langle\vec{d}\right|\rho_{3}\left|\vec{d}\right\rangle\right|\\=&\sum_{\{\{m\}|m\geq2\}}\mathcal{P}_{\{m\}}\epsilon|\langle\vec{d}|\xi_{\{m\}}\rangle|^2+\mathcal{P}_{\{m-1\}}\left(1-\epsilon\right)|\langle\vec{d}|\xi_{\{m-1\}}^\prime\rangle|^2\\=&\epsilon\left(\mathcal{A}^\prime_{\vec{d}}+\mathcal{B}^\prime_{\vec{d}}\cos\phi\right)+\mathcal{C}^\prime_{\vec{d}},
    \end{aligned}
\end{equation}
which is in the same form as Eq.\eqref{equ:Pd_propto}. For the simplest virtual outcome $\vec{d}$ that only detectors $D_{a,k}$ and $D_{b,l}$ click, we get
\begin{equation}\label{equ:heralded_result_2}
    \begin{aligned}
&\left|\left\langle\vec{d}|\xi^\prime_{\{m\}}\right\rangle\right|^2=\left|A_{\{m\}}\mathrm{e}^{i\phi}+B_{\{m\}}\right|^2,\\
        A_{\{m\}}=&\sum_{\{n\}}\sqrt{\frac{(\sum_{i}n_i+1)!(\sum_{i}\left(m_i-n_i\right))!}{2^{\sum_{i}m_i+1}}}U_{1k}\\
        &\cdot\prod_{i=2}^{N}\left(U_{ik}^{n_i}U_{il}^{m_i-n_i}\binom{m_i}{n_i}\right),\\
   B_{\{m\}}=&\sum_{\{n\}}\sqrt{\frac{(\sum_{i}n_i)!(\sum_{i}\left(m_i-n_i\right)+1)!}{2^{\sum_{i}m_i+1}}}U_{1l}\\
        &\cdot\prod_{i=2}^{N}\left(U_{ik}^{n_i}U_{il}^{m_i-n_i}\binom{m_i}{n_i}\right);
    \end{aligned}
\end{equation}
and
\begin{equation}\label{equ:heralded_result_3}
    \begin{aligned}
        &\left|\left\langle \vec{d}|\xi_{\{m\}}\right\rangle\right|^2=|C_{\{m\}}|^2\\=&\left|\sum_{\{n\}}\sqrt{\frac{(\sum_{i}n_i)!(\sum_{i}\left(m_i-n_i\right))!}{2^{\sum_{i}m_i}}}\prod_{i=2}^{N}\left(U_{ik}^{n_i}U_{il}^{m_i-n_i}\binom{m_i}{n_i}\right)\right|^2,
    \end{aligned}
\end{equation}
where $\{n\}=\sum_{n_i}\{n_1,\cdots,m_N|n_i\leq m_i\}$ denotes the number of photons emitted by each source $S_i$ and then collected by the detector $D_{a,k}$.
The values of $\mathcal{A}^\prime_{\vec{d}},\mathcal{B}^\prime_{\vec{d}}\; \text{and}\;\mathcal{C}^\prime_{\vec{d}}$ in Eq.\eqref{equ:heralded_result_1} are
\begin{equation}\label{equ:herald_result_important2}
    \begin{aligned}
        \mathcal{A}^\prime_{\vec{d}}&= \mathcal{P}_{\{m-1\}}\epsilon\left(|A_{\{m-1\}}|^2+|B_{\{m-1\}}|^2\right)-\mathcal{P}_{\{m\}}\left(1-\epsilon\right)|C_{\{m\}}|^2,\\
        \mathcal{B}^\prime_{\vec{d}}&= 2\mathcal{P}_{\{m-1\}}\epsilon|A_{\{m-1\}}||B_{\{m-1\}}|,\\
        \mathcal{C}^\prime_{\vec{d}}&= \mathcal{P}_{\{m\}}\left(1-\epsilon\right)|C_{\{m\}}|^2.
    \end{aligned}
\end{equation}
Given Eqs. \eqref{equ:heralded_result_1}-\eqref{equ:herald_result_important2}, the values of $\{\mathcal{P}(\{m\}|\phi)\}$ in Eq.(21) of the main text can be numerically calculated.

\subsection{CLI quantum-enhanced interferometer with imperfect detectors}
We have also performed numerical simulations of our scheme under imperfect detector conditions. Denoting the detector efficiency by $\xi$ and the dark-count probability by $p_d$, we denote
$P\left(\vec{d}\right)$ as the probability of each measurement outcome $\vec{d}$ of the virtual method with perfect detector responses,
and $P^{\prime}(\vec{d})$ as the probability of the same measurement outcome $\vec{d}$ of the virtual method when accounting for finite detector efficiency $\xi<1$. 
Under the virtual method with finite detector efficiency $\xi$, any recorded detector outcome not only arises from the exact set of detectors that genuinely clicked, but also contains cases in which additional detectors should have clicked but remain silent:
\begin{equation}
    P^\prime{(\vec{d})}=\sum_{\vec{d} \in \{\vec{d}^\prime\}}\xi^{n}(1-\xi)^{n^\prime-n}P{(\vec{d}^\prime)}.
\end{equation}
Here, $n$ denotes the number of detectors that actually clicked, $n'$ is the number of detectors that would have clicked under perfect $(\xi=1)$ efficiency, and $\{\vec{d}'\}$ represents the set of all ideal measurement results that can produce the observed outcome $\vec{d}$ through imperfect detectors.

When both the detector efficiency $\xi$ and the channel loss $p_{d}$ are taken into account, the probability expressions become more complex. As an example, we consider the case that detectors $D_{a,i}\;\text{and}\;D_{b,j}$ both click:
\begin{align*}
    P^{\prime\prime}(D_{a,i},D_{b,j})=&p_d^2(1-p_d)^2P(\text{null})+p_d(1-p_d)^2(P^\prime(D_{a,i})\\+&P^\prime(D_{b,j}))1-p_d)^2P^{\prime}(D_{a,i},D_{b,j}).
\end{align*}

\begin{figure}
    \centering
    \includegraphics[width=0.5\linewidth]{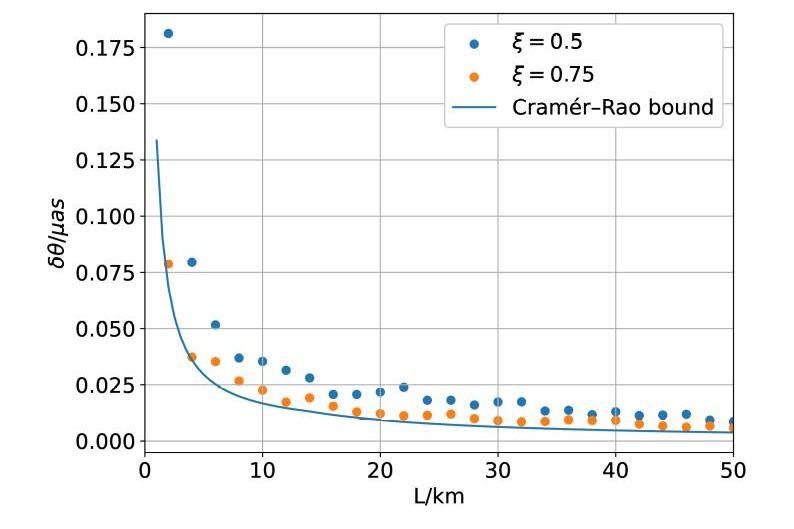}
    \caption{Average angular error versus baseline length for various detector efficiencies, incorporating detector efficiency and dark-count effects, at a fixed number of $N_t=10^8$ time windows. For each simulation data point, several hundred random angle values $\theta$ uniformly distributed in $[0, \lambda/L]$ are generated.}
    \label{fig:enter-label4}
\end{figure}
As shown in Fig. \ref{fig:enter-label4}, we simulate the average angular uncertainty of our method with imperfect threshold detectors under different baseline lengths. With a $10\;\text{km}$ long baseline and a detector efficiency of $50 \%$, our method can achieve an angular uncertainty below 0.05 $\mu as$ using $10^8$ time windows, or reach 0.01 $\mu as$ under the same number of time windows and $50\;\mathrm{km}$ baseline length, thus demonstrating its experimental feasibility under practical detectors.
\end{document}